\documentclass[a4paper,12pt]{article}
\usepackage[margin=1 in]{geometry}
\usepackage{amsmath}
\usepackage{amsfonts}
\usepackage{amssymb}
\usepackage[all]{xy}           %xypic macro for latex2.09
\usepackage[normalem]{ulem}
\usepackage{graphicx} % Figures
\usepackage{fancyhdr}
\usepackage{longtable} % Multi-page tables
\usepackage{bbding}
\usepackage{txfonts}
\usepackage{comment}
\usepackage[shortlabels]{enumitem}
\usepackage{ifpdf}
\ifpdf
 \usepackage[final,backref=page,hyperindex]{}
\else
 \usepackage[final,backref=page,hyperindex,hypertex]{}
\fi
\usepackage{tikz}
\usepackage{xspace}
\usepackage{authblk}
\usepackage{lineno}
\usepackage{multirow}
\usepackage{xcolor}
\usepackage{float}
\usepackage{rotating}

\usepackage[nodisplayskipstretch]{setspace}
\usepackage[comma,super,sort&compress]{natbib}

\title{Adaptive dose-response studies to establish proof-of-concept in learning-phase clinical trials}

\author{Shiyang Ma$^{1,\ast}$, Michael P. McDermott$^2$}
\date{}                     %% if you don't need date to appear
\providecommand{\keywords}[1]{{KEY WORDS: } #1}

\begin{document}

%\pagewiselinenumbers
\maketitle

\noindent
{$^1$ Department of Biostatistics, Columbia University, New York, NY, 10032, USA}\\
{$^2$ Department of Biostatistics and Computational Biology, University of Rochester, Rochester, NY, 14642, USA}\\
{$^\ast$ e-mail: sm4857@columbia.edu}\\

\begin{abstract}
In learning-phase clinical trials in drug development, adaptive designs can be efficient and highly informative when used appropriately. In this article, we extend the multiple comparison procedures with modeling techniques (MCP-Mod) procedure with generalized multiple contrast tests (GMCTs) to two-stage adaptive designs for establishing proof-of-concept. The results of an interim analysis of first-stage data are used to adapt the candidate dose-response models and the dosages studied in the second stage. GMCTs are used in both stages to obtain stage-wise $p$-values, which are then combined to determine an overall $p$-value. An alternative approach is also considered that combines the $t$-statistics across stages, employing the conditional rejection probability (CRP) principle to preserve the Type I error probability. Simulation studies demonstrate that the adaptive designs are advantageous compared to the corresponding tests in a non-adaptive design if the selection of the candidate set of dose-response models is not well informed by evidence from preclinical and early-phase studies.\end{abstract}


\label{firstpage}
%\doublespacing
\setstretch{1.7} %25 line per page

%\newpage

\section{Introduction}

Motivated by  the desire for greater efficiency in drug development and the low success rates in confirmatory (Phase 3) studies, methodological research on adaptive designs and interest in their application has grown tremendously over the last 30 years. In an adaptive design, accumulating data can be used to modify the course of the trial. Several possible adaptations can be considered in interim analyses, for example, adaptive randomization for dose finding, dropping and/or adding treatment arms, sample size re-estimation, and early stopping for safety, futility or efficacy, to name a few.

Validity and integrity are two major considerations in adaptive designs (Dragalin, 2006). Because data from one stage of the trial can inform the design of future stages of the trial, careful steps need to be taken to maintain the validity of the trial, i.e., control of the Type I error probability and minimization of bias. To maintain trial integrity, it is important that all adaptations be pre-planned, prior to the unblinded examination of data, and that all trial personnel other than those responsible for making the adaptations are blind to the results of any interim analysis (Food and Drug Administration, 2019). It is also important to ensure consistency in trial conduct among the different stages.

A general method for hypothesis testing in experiments with adaptive interim analyses based on combining stage-wise $p$-values was proposed by Bauer and K\"{o}hne (1994). The basic idea behind the construction of a combination test in a two-stage adaptive design is to transform the stage-wise test statistics to $p$-values, with independence of the $p$-values following from the conditional invariance principle (Brannath {\it{et al}}., 2007, 2012; Wassmer and Brannath, 2016), regardless of the adaptation performed after the first stage. The principle holds as long as the null distribution of the first-stage $p$-value ($p_1$) as well as the conditional distribution of the second-stage $p$-value ($p_2$) given $p_1$ are stochastically larger than the $U(0,1)$ distribution (the so-called ``p-clud" property). A specified combination function is used to combine the $p$-values obtained before and after the preplanned adaptation of the design into a single global test statistic. An extension of combination tests to allow more flexibility regarding the number of stages and the choice of decision boundaries was provided by Brannath {\it{et al}}. (2002).

In dose-response studies, a component of the MCP-Mod procedure (Bretz {\it{et al}}., 2005) has gained popularity for the purpose of detecting a proof-of-concept (PoC) signal in learning-phase trials. The procedure consists of specifying a set of candidate dose-response models, determining the optimal contrast statistic for each candidate model, and using the maximum contrast as the overall test statistic. Other authors have considered extensions of this procedure to adaptive dose-response designs. Miller (2010) investigated a two-stage adaptive dose-response design for PoC testing incorporating adaptation of the dosages, and possibly the contrast vectors. He developed an adaptive multiple contrast test (AMCT) that combines the multiple contrast test statistics across two stages under the assumption that the variance is known. Franchetti {\it{et al}}. (2013) extended the MCP-Mod procedure to a two-stage dose-response design with a pre-specified rule of adding and/or dropping dosage groups in Stage 2 based on the Stage 1 results. The PoC test uses Fisher's (1932) combination method to combine the two stage-wise $p$-values, each obtained by applying the MCP-Mod procedure to the data from each stage. This method includes a restrictive requirement of equal total sample sizes for each stage. Also, the authors claimed that the independence of the two stage-wise $p$-values is potentially compromised if the number of dosages used in Stage 2 is not the same as that used in Stage 1 and proposed a method for assigning weights to the different dosage groups to deal with this problem. We do not believe that such weighting is necessary as long as the statistic used to combine the stage-wise $p$-values (Fisher's, in this case) does not include weights that depend on the Stage 1 data.

Early work related to adaptive designs for dose-response testing includes a general procedure with multi-stage designs proposed by Bauer and R\"{o}hmel (1995), in which dosage adaptations were performed at interim analyses. Other goals of adaptive dose-response studies include determining if any dosage yields a clinically relevant benefit, estimating the dose-response curve, and selecting a target dosage for further study (Dragalin {\it{et al}}., 2010). Several model-based adaptive dose-ranging designs that utilize principles of optimal experimental design to address these objectives were studied by Dragalin {\it{et al}}. (2010). Bornkamp {\it{et al}}. (2011) proposed a response-adaptive dose-finding design under model uncertainty, which uses a Bayesian approach to update the parameters of the candidate dose-response models and model probabilities at each interim analysis.

In this article, we propose new methods to address the specific objective of detecting a PoC signal in adaptive dose-response studies with normally-distributed outcomes. We extend the MCP-Mod procedure to include generalized multiple contrast tests (GMCTs; Ma and McDermott, 2020) and apply them to adaptive designs; we refer to these as adaptive generalized multiple contrast tests (AGMCTs). These tests are introduced in Section \ref{Adaptive Generalized Multiple Contrast Tests}. In Section \ref{Adaptive Multiple Contrast Test} we extend the AMCT of Miller (2010) to accommodate more flexible adaptations and to the important case where the variance is unknown using the conditional rejection probability (CRP) principle (M\"{u}ller and Sch\"{a}fer, 2001, 2004). Numerical examples are provided in Section \ref{Numerical Example} to illustrate the application of the AGMCTs and AMCT. In Section \ref{Simulation studies}, we conduct simulation studies to evaluate the operating characteristics of the various methods as well as the corresponding tests for non-adaptive designs. The conclusions are given in Section \ref{Conclusion}.

\section{Adaptive Generalized Multiple Contrast Tests}\label{Adaptive Generalized Multiple Contrast Tests}

In this section, we propose a two-stage adaptive design in which we use data from Stage 1 to get a better sense of the true dose-response model and make adaptations to the design for Stage 2. We then use data from both Stage 1 and Stage 2 to perform an overall test to detect the PoC signal. The rationale is to overcome the problem of potential model misspecification at the design stage.

\subsection{General Procedure}
We consider the case of a normally distributed outcome variable. Suppose that there are $n_{i1}$ subjects in dosage group $i$ in Stage 1, $i=1,\ldots,k_1$. Denote the first stage data as $\pmb{Y}_1=(Y_{111},\ldots,Y_{1 n_{11} 1},\ldots, $ $Y_{k_1 1 1},\ldots,Y_{k_1 n_{k_1 1}1})^\prime.$
The statistical model is
$$Y_{ij1}=\mu_i +\epsilon_{ij1},\quad\epsilon_{ij1}\stackrel{iid}{\sim} N(0, \sigma^2),\quad i=1,\ldots, k_1,\ j=1,\ldots, n_{i1}.$$
The true mean configuration is postulated to follow some dose-response model $\mu_i=f(d_i,\pmb{\theta})$, where $d_i$ is the dosage in the $i^{\text{th}}$ group, $i=1,\ldots,k_1$. The dose-response model is restricted to be of the form $f(\cdot;\pmb{\theta})=\theta_0+\theta_1 f^0(\cdot;\pmb{\theta}^0)$, where $f^0(\cdot;\pmb{\theta}^0)$ is a standardized dose-response model indexed by a parameter vector $\pmb{\theta}^0$ (Thomas, 2017). A candidate set of $M$ dose-response models $f_m(\cdot,\pmb{\theta})$, $m=1,\ldots,M$, including values for $\pmb{\theta}$, is pre-specified. For each candidate model, an optimal contrast is determined to maximize the power to detect differences among the mean responses; the contrast coefficients are chosen to be perfectly correlated with the mean responses if that model is correct (Bretz {\it{et al}}., 2005; Pinheiro {\it{et al}}., 2014).

For each candidate model, the following hypothesis is tested:
$$H_{0m 1}: \sum_{i=1}^{k_1} c_{mi1} \mu_i=0,\quad\text{vs.}\quad H_{1m 1}: \sum_{i=1}^{k_1} c_{mi1} \mu_i>0,\quad m=1,\ldots, M,$$
where $c_{m11},\ldots, c_{mk_11}$ are the optimal contrast coefficients associated with the $m^{\text{th}}$ candidate model in Stage 1. The multiple contrast test statistics are
$$T_{m1}=\sum_{i=1}^{k_1} c_{mi1} \bar{Y}_{i1}\Bigg/\left(S_1\sqrt{\sum_{i=1}^{k_1}\frac{c_{mi1}^2}{n_{i1}}}\right),\quad m=1,\ldots, M,$$
where $\bar{Y}_{i1}=\sum_{j=1}^{n_{i1}} Y_{ij1}/n_{i1}$ and the pooled variance estimator is $S_1^2=\sum_{i=1}^{k_1}\sum_{j=1}^{n_{i1}} (Y_{ij1}-\bar{Y}_{i1})^2/\nu_1$, where $\nu_1=\sum_{i=1}^{k_1} n_{i1}-k_1$. The joint null distribution of $(T_{11},\ldots,T_{M1})^\prime$ is multivariate $t$ (with $\nu_1$ degrees of freedom) with common denominator and correlation matrix having elements
$$\rho_{m m^{\prime}1}=\sum_{i=1}^{k_1}\frac{c_{mi1} c_{m^{\prime} i1}}{n_{i1}}\Bigg/\sqrt{\sum_{i=1}^{k_1}\frac{c_{m i1}^2}{n_{i1}}\sum_{i=1}^{k_1}\frac{c_{m^{\prime} i1}^2}{n_{i1}}}, \quad m, m^{\prime}=1,\ldots, M.$$

Let $p_{m1}=1-\mathcal{T}_{\nu_1} (T_{m1})$ be the $p$-values derived from $T_{m1}$, $m=1, \ldots, M$, where $\mathcal{T}_{\nu_1} (\cdot)$ is the cumulative distribution function of the $t$ distribution with $\nu_1$ degrees of freedom. We consider three combination statistics to combine the $M$ dependent one-sided $p$-values in Stage 1 (Ma and McDermott, 2020):
\begin{enumerate}[(i)]
\item Tippett's (1931) combination statistic,
$$\Psi_{T1}=\min_{1 \leq m \leq M} \ p_{m1};$$
\item Fisher's (1932) combination statistic,
$$\Psi_{F1}=-2 \sum_{m=1}^M \log (p_{m1});$$
\item Inverse normal combination statistic (Stouffer, 1949),
$$\Psi_{N1}=\sum_{m=1}^M \Phi^{-1} (1-p_{m1}).$$
\end{enumerate}

%\noindent Bullet lists are not allowed. Always use (i), (ii), etc.

Note that the use of Tippett's combination statistic is equivalent to the original MCP-Mod procedure; the use of different combination statistics results in a generalization of the MCP-Mod procedure, yielding GMCTs (Ma and McDermott, 2020). When the $p$-values are independent, these statistics have simple null distributions. In our case the $p$-values are dependent, but the correlations among $T_{11},\ldots,T_{M1}$ are known. For Tippett's combination method, one can obtain multiplicity-adjusted $p$-values from $T_{m1}$, $m=1, \ldots, M$, given the correlation structure using the {\tt{mvtnorm}} package in {\tt{R}}. A PoC signal is established in Stage 1 if the minimum adjusted $p$-value $p_{\text{min, adj}1}< \alpha$ (Bretz {\it{et al}}., 2005). For Fisher's and the inverse normal combination methods, excellent approximations to the null distributions of $\Psi_{F1}$ and $\Psi_{N1}$ have been developed (Kost and McDermott, 2002), enabling computation of the overall $p$-value $p_1$ for Stage 1 using a GMCT (Ma and McDermott, 2020).

After obtaining the Stage 1 data, we make design adaptations and determine the optimal contrasts for the updated models in Stage 2 (see Sections \ref{Adapting the Candidate Dose-Response Models} and \ref{Adapting the Dosage Groups} below). We then conduct a GMCT in Stage 2 and obtain the second-stage $p$-value $p_2$. Under the overall null hypothesis $H_0: \mu_1= \cdots =\mu_{k^*}$, where $k^*$ is the total number of unique dosage groups in Stages 1 and 2 combined, the independence of the stage-wise $p$-values $p_1$ and $p_2$ can be established using the conditional invariance principle (Brannath {\it{et al}}., 2007). To perform the overall PoC test in the two-stage adaptive design, we combine $p_1$ and $p_2$ using one of the above combination statistics.

A procedure that ignores the adaptation, i.e., that simply pools the data from Stage 1 and Stage 2 and applies a GMCT to the pooled data as if no adaptation had been performed, would substantially increase the Type I error probability.

\subsection{Adapting the Candidate Dose-Response Models}\label{Adapting the Candidate Dose-Response Models}
Here and in Section \ref{Adapting the Dosage Groups} below, we consider adaptations for the second stage that are arguably most relevant for PoC testing, namely those of the candidate dose-response models and the dosages to be studied. The choice of the candidate dose-response models and dosages for Stage 1 would depend on prior knowledge from pre-clinical or early-stage clinical experience with the investigative agent. If there is great uncertainty concerning the nature of the dose-response relationship, it would seem sensible to select a more diverse set of candidate dose-response models with pre-specified parameters when the trial begins.

After collecting the Stage 1 data, these data can be used to estimate $\pmb{\theta}$ for each of the $M$ candidate dose-response models and adapt each of the models by substituting $\pmb{\hat{\theta}}$ for the original specification (guess) of $\pmb{\theta}$. The optimal contrast vectors can be constructed for each of the updated models $f_m(\cdot,\pmb{\hat{\theta}})$, $m=1,\ldots,M$, for use in Stage 2.

%\textbf{MODIFY THE WAY TO HANDLE THE CASE OF NON-CONVERGENCE FOR NONLINEAR MODEL}

A potential problem occurs when the true dose-response model differs markedly from some of the specified candidate models and if those candidate models are nonlinear models with several unknown parameters. In such cases there can be a failure to fit the models using the Stage 1 data. To handle this problem, one can consider fall-back approaches to determine the corresponding contrasts to be used in Stage 2. These include using isotonic regression (Robertson {\it{et al}}., 1988), imposing reasonable bounds on the nonlinear parameters during model-fitting (as is done in the {\tt{R}}-package {\tt{DoseFinding}} to ensure the existence of the maximum likelihood estimates), and retaining the Stage 1 contrast for use in Stage 2. Different strategies can be used for different models in cases where more than one model cannot be fit using the Stage 1 data.

Specifically, consider the following 5 candidate dose-response models:
\begin{itemize}
  \item[] $E_{\max}$ model: $f_1(d,\pmb{\theta})=E_0+E_{\max} d/ (ED_{50}+d)$
  \item[] Linear-log model: $f_2 (d,\pmb{\theta})=\theta_0+\theta_1\log(5d+1)$
  \item[] Linear model: $f_3 (d,\pmb{\theta})=\theta_0+\theta_1 d$
  \item[] Quadratic model: $f_4 (d,\pmb{\theta})=\theta_0+\theta_1 d + \theta_2 d^2$
  \item[] Logistic model: $f_5 (d,\pmb{\theta})=E_0+E_{\max} /[1 + \exp\{(ED_{50}-d)/ \delta\}]$
\end{itemize}
Among these 5 candidate models, the $E_{\max}$ and Logistic models are the ones that may fail to converge since the others can be expressed as linear models in $d$ (or a simple function of $d$). A possible fall-back strategy could be as follows: if only one of the $E_{\max}$ and Logistic models fails to converge in Stage 1, isotonic regression is used to generate the corresponding contrast for use in Stage 2; if both the $E_{\max}$ and Logistic models fail to converge in Stage 1, then isotonic regression is used to generate the corresponding contrast for the Logistic model and the same contrast that was used in Stage 1 is used in Stage 2 for the $E_{\max}$ model (see Section \ref{Numerical Example, AGMCT} for a numerical example).

Another potential concern arises if the data from Stage 1 suggest that there is a negative dose-response relationship, i.e., that higher dosages are associated with worse outcomes. In this case, the adapted contrast associated with the linear model, say, in Stage 2 would be the negative of that used in Stage 1. If a similar dose-response pattern is observed in Stage 2, then the contrast associated with the linear model would incorrectly indicate (possibly strong) evidence against the null hypothesis. One way to avoid this problem would be to not adapt the dose-response models in such a case, but instead to consider adapting the dosage groups by retaining only dosages, if any, that appear to be associated with increasing sample means (see Section \ref{Adapting the Dosage Groups} below).

Ideally, of course, it would be required to pre-specify the measures that would be taken to deal with the problems noted above (non-convergence of non-linear models, negative dose-response relationship) prior to examination of the data.

One could also consider different numbers of candidate models (or contrast vectors) in Stage 1 and Stage 2. One non-model-based option, for example, would be to use a single contrast in Stage 2 based on the sample means of the dosage groups from Stage 1. We found that this strategy, while intuitively appealing, yielded tests with reduced power, likely due to the reliance on a single contrast combined with the uncertainty associated with estimation of the means of each dosage group in Stage 1. One could also consider a small number of other contrasts based on values that are within the bounds of uncertainty reflected in the sample means, though how to choose these contrasts is somewhat arbitrary.

\subsection{Adapting the Dosage Groups}\label{Adapting the Dosage Groups}

Adaptation of the dosage groups in Stage 2, including the number of dosage groups, could also be considered. One would have to establish principles for adding and/or dropping dosages; for example, dropping active dosages that appear to be less efficacious than placebo or that appear to be less efficacious than other active dosages, or adding a dosage (within a safe range) when there appears to be no indication of a dose-response relationship in Stage 1. Relevant discussion of these issues can be found in Bauer and R\"{o}hmel (1995), Miller (2010), and Franchetti {\it{et al}}. (2013).

To illustrate this type of adaptation, we create an example dosage adaptation rule to drop the active dosage groups that appear to be less efficacious than placebo and the adjacent group. Suppose that there are $k_1$ dosage groups in Stage 1 and denote the dosage vector in Stage 1 as $\pmb{d}_{\text{Stage1}}=(d_{11},\ldots, d_{k_11})^\prime$, where $d_{11}=0$ (placebo group). We will select $k_2$ dosage groups from the $k_1$ Stage 1 dosage groups, $k_2\leq k_1$. Denote the dosage vector in Stage 2 as $\pmb{d}_{\text{Stage2}}=(d_{12}, \ldots, d_{k_22})^\prime$, where $d_{12}=0$ (placebo group). The example dosage adaptation rule is as follows:
\begin{itemize}
  \item[] \textbf{Step 1}: Always select the placebo group to be included in Stage 2, i.e., $d_{12}=d_{11}=0$.

  \item[] \textbf{Step 2}: Consider the difference in the means between each active dosage group and the placebo group in Stage 1.

Denote $\hat{\Delta}_{21}=\bar{Y}_{21}-\bar{Y}_{11},\ldots,\hat{\Delta}_{k_11}=\bar{Y}_{k_11}-\bar{Y}_{11}$. If there exists dosage group(s) $i$, $i=2,\ldots, k_1$, such that $\hat{\Delta}_{i 1}<-\delta$, where $\delta\ge 0$, then we remove dosage(s) $d_{i 1}$ from consideration; however, if $\hat{\Delta}_{i 1}<-\delta$ for all $i=2,\ldots, k_1$, then we stop the trial at the interim analysis and fail to reject $H_0$.

  \item[] \textbf{Step 3}: Consider the differences in the means between two adjacent dosage groups among the remaining dosage groups, ordered from smallest to largest.

After Steps 1 and 2, we have selected $d_{11}$ (placebo) into Stage 2 and have several remaining dosage groups $d_{\tilde{2}1},\ldots,d_{\tilde{k}1}$, where $\tilde{k}\leq k_1$.

We first examine the difference in the means between dosages $d_{11}$ and $d_{\tilde{2}1}$. If $\hat{\Delta}_{\tilde{2}1}=\bar{Y}_{\tilde{2}1}-\bar{Y}_{11} > -\delta$, then $d_{\tilde{2}1}$ is selected to be included in Stage 2, i.e., $d_{22}=d_{\tilde{2}1}$; otherwise, $d_{\tilde{2}1}$ is discarded and we proceed to the next possible dosage $d_{\tilde{3}1}$.

If $d_{\tilde{2}1}$ is selected to be included in Stage 2, then we proceed to compare the means between dosages $d_{\tilde{2}1}$ and $d_{\tilde{3}1}$. If $\hat{\Delta}_{\tilde{3} \tilde{2}}=\bar{Y}_{\tilde{3}1}-\bar{Y}_{\tilde{2}1}> -\delta$, then $d_{\tilde{3}1}$ is selected to be included in Stage 2, i.e., $d_{32}=d_{\tilde{3}1}$; otherwise, $d_{\tilde{3}1}$ is discarded. However, if $d_{\tilde{2}1}$ is discarded, then the means should be compared between dosages $d_{11}$ and $d_{\tilde{3}1}$, since these are now adjacent dosages among those remaining.

This procedure is repeated until the last possible dosage $d_{\tilde{k}1}$ is reached and its associated mean is compared with that of the remaining adjacent dosage. This results in a final number $k_2 \leq \tilde{k}$ of dosage groups selected to be included in Stage 2, i.e., $\pmb{d}_{\text{Stage2}}=(d_{12}, \ldots, d_{k_22})^\prime$.
\end{itemize}

Here we consider the threshold of adaptive dosing $\delta=0$, which simply considers the difference between two sample means and retains the dosage with the larger sample mean. This threshold might be strict since it does not consider the variability of the difference between two sample means. An alternative threshold could be $\delta=\sqrt{\text{var}(\bar{Y}_{i1}-\bar{Y}_{i^\prime 1})}$, $i,i^\prime=1, \ldots, k_1$, which retains a dosage with a mean that is no more than one standard error lower than the mean of the adjacent dosage (or placebo). Users are free to choose their own threshold $\delta$ based on considerations specific to their problem.

%$\hat{\Delta}_{21}=\bar{Y}_{21}-\bar{Y}_{11}>-\delta \Rightarrow \bar{Y}_{21}>\bar{Y}_{11}-\delta=\bar{Y}_{11}-\sqrt{\text{var}(\bar{Y}_{21}-\bar{Y}_{11})}$.

%$\bar{Y}_{21}>\bar{Y}_{11}-(\bar{Y}_{21}-\bar{Y}_{11})/ \sqrt{\text{var}(\bar{Y}_{21}-\bar{Y}_{11})}$

We emphasize that this is just one possible rule to adapt the dosage groups for Stage 2, and this rule only considers dropping dosages at the end of Stage 1. One could consider different adaptation rules that allow adding and/or dropping dosages at the end of Stage 1, i.e., $k_2$ does not need to be less than or equal to $k_1$, and some of the dosage groups selected in Stage 2 may differ from those included in Stage 1. Also, as in Miller (2010), such a rule is based on heuristic considerations and is relatively easy to communicate to non-statisticians. Mercier {\it{et al}}. (2015) provide an approach to selecting dosages for Stage 2 based on the hypothetical dose-response shape (out of several pre-specified) that correlates highest with the data observed in Stage 1.

One can adapt both the candidate dose-response models and the dosage groups in Stage 2. The optimal contrast vectors for Stage 2 would then be determined by the updated candidate dose-response models with parameters $\pmb{\hat{\theta}}$ and the adapted dosages $\pmb{d}_{\text{Stage2}}$. The overall p-value for Stage 2, $p_2$, would be obtained from a GMCT that uses the updated optimal contrast vectors. We incorporate this strategy in our simulation studies below. It should be noted that if one adapts only the candidate dose-response models and not the dosage groups, the contrasts for the Linear and Linear-log models would not change based on the Stage 1 data.  This would not be the case if one also adapted the dosage groups.

\section{Adaptive Multiple Contrast Test}\label{Adaptive Multiple Contrast Test}

\subsection{Known Variance Case}\label{AMCT, Known Variance Case}

Instead of combining the stage-wise $p$-values $p_1$ and $p_2$, each based on a GMCT, Miller (2010) suggested combining the test statistics for each candidate dose-response model across the two stages, and then derving an overall $p$-value from a multiple contrast test applied to those statistics, assuming a known variance $\sigma^2$. For each candidate model, we have
$$Z_m=\left(\sum_{i=1}^{k_1} c_{mi1} \bar{Y}_{i1}+\sum_{i=1}^{k_2} c_{mi2} \bar{Y}_{i2}\right)\Bigg/\sigma\sqrt{\sum_{i=1}^{k_1} \frac{c^2_{mi1}}{n_{i1}}+\sum_{i=1}^{k_2}\frac{c^2_{mi2}}{n_{i2}}},\quad m= 1,\ldots,M.$$
Since $k_2$, $c_{mi2}$, and $n_{i2}$ can depend on the interim data (adaptation), the null distribution of $Z_m$ is not standard normal in general.

In order to control the Type I error probability of the overall test, Miller (2010) applies a conditional error approach based on the conditional rejection probability (CRP) principle (M\"{u}ller and Sch\"{a}fer, 2001, 2004). Computation of the conditional Type I error probability requires pre-specification of what Miller (2010) calls a ``base test", i.e., pre-specified values for the contrast coefficients ($c^*_{mi2}$), number of dosage groups ($k_2^*$), and group sample sizes ($n^*_{i2}$) in Stage 2, $i=1, \ldots, k^*_2$, $m=1,\ldots, M$. There is not a clear best strategy for choosing these pre-specified values. Miller (2010) considers an example where all possible Stage 2 designs can be enumerated and have $k_1=k_2$ and $n_{i1}=n_{i2}$, $i=1,\ldots,k_1$, and the pre-specified values involving $c_{mi2}^*$, $i=1,\ldots,k_2$, $m=1,\ldots,M$, are averaged over the possible Stage 2 designs. More generally one cannot enumerate all possible Stage 2 designs, so in the development below we pre-specify $c^*_{mi2}=c_{mi1}$, $k^*_2=k_1$, and $n^*_{i2}=n_{i1}$, $i=1,\ldots,k_2$, $m=1,\ldots,M$. Since the dosages can also be adapted, we suggest pre-specifying $\pmb{d}^*_{\text{Stage2}}=\pmb{d}_{\text{Stage1}}=(d_{11},\ldots,d_{k_1 1})^\prime$. One can think of this ``base test" as one that is based on a study that uses the same design in Stage 2 as was used in Stage 1.

The $Z$-statistics for the base test are
$$Z^*_m=\sum_{i=1}^{k_1} c_{mi1}\left(\bar{Y}_{i1}+\bar{Y}_{i2}\right)\Bigg/\sigma\sqrt{2\sum_{i=1}^{k_1}\frac{c_{mi1}^2 }{n_{i1}}},\quad m=1,\ldots,M.$$
Under $H_0$, the joint distribution of $\pmb{Z}^*=(Z^*_1,\ldots,Z^*_M)^\prime$ is multivariate normal with mean $\pmb{0}$ and covariance matrix $\pmb{R}^*=(\rho_{m m^{\prime}1})$, $m, m^{\prime}=1,\ldots, M$. One can then obtain the non-adaptive $\alpha$-level critical value $u^*_{1-\alpha}$ based on the null distribution of $Z_{\max}^*=\max \{\pmb{Z}^*\}$ using the {\tt{R}}-package {\tt{mvtnorm}}.

In order to obtain the conditional Type I error probability $A=P_{H_0}(Z^*_{\max}\ge u^*_{1-\alpha}\,|\,\pmb{Y}_1)$, where $\pmb{Y}_1$ are the Stage 1 data, it can be seen that the conditional distribution of $\pmb{Z}^*$ given $\pmb{Y}_1=\pmb{y}_1$ is multivariate normal with mean vector
$$\left(\sum_{i=1}^{k_1} c_{1i1}\bar{y}_{i1}\Bigg/\sigma\sqrt{2\sum_{i=1}^{k_1}\frac{c_{1i1}^2}{n_{i1}}},\ldots,\sum_{i=1}^{k_1} c_{Mi1}\bar{y}_{i1}\Bigg/\sigma\sqrt{2\sum_{i=1}^{k_1}\frac{c_{Mi1}^2 }{n_{i1}}}\right)^\prime$$
and covariance matrix $\pmb{R_2}^*=\pmb{R}^*/2$, where $\bar{y}_{i1}=\sum_{j=1}^{n_{i1}} y_{ij1}/n_{i1}$, $i=1,\ldots,k_1$. Hence, the conditional Type I error probability is
$$A=P_{H_0} (Z^*_{\max}\geq u^*_{1-\alpha}\,|\,\pmb{Y}_1)=1- P_{H_0} (\pmb{Z}^* \leq (u^*_{1-\alpha},\ldots,u^*_{1-\alpha})^\prime\,|\,\pmb{Y}_1),$$
which can be obtained using the {\tt{pmvnorm}} function in the {\tt{R}}-package {\tt{mvtnorm}}.

In general, the interim analysis at the end of Stage 1 could yield adapted values of $c_{mi2}$, $k_2$, and $n_{i2}$ for Stage 2 and, hence, the adapted $Z$-statistics $Z_m$, $m=1, \ldots, M$. Denote $\pmb{Z}=(Z_1,\ldots,Z_M)^\prime$ and $Z_{\max}=\max \{\pmb{Z}\}$. The adaptive critical value $\tilde{u}_{1-\alpha}$ can be obtained by solving the equation
$$\tilde{A}=P_{H_0}(Z_{\max}\geq\tilde{u}_{1-\alpha}\,|\,\pmb{Y}_1)=1- P_{H_0}(\pmb{Z} \leq (\tilde{u}_{1-\alpha},\ldots,\tilde{u}_{1-\alpha})^\prime\,|\,\pmb{Y}_1)= A,$$
where the conditional distribution of $\pmb{Z}$ given $\pmb{Y}_1$ is multivariate normal with mean vector
$$\left(\sum_{i=1}^{k_1} c_{1i1}\bar{y}_{i1}\Bigg/\sigma\sqrt{\sum_{i=1}^{k_1}\frac{c_{1i1}^2}{n_{i1}}+\sum_{i=1}^{k_2}\frac{c_{1i2}^2}{n_{i2}}},\ldots,\sum_{i=1}^{k_1} c_{Mi1}\bar{y}_{i1}\Bigg/\sigma\sqrt{\sum_{i=1}^{k_1}\frac{c_{Mi1}^2}{n_{i1}}+\sum_{i=1}^{k_2}\frac{c_{Mi2}^2}{n_{i2}}}\right)^{\prime}$$
and covariance matrix $\pmb{\tilde{R}}=(\text{cov}(Z_m,Z_{m^\prime}\,|\,\pmb{Y}_1))$, $m,m^\prime=1,\ldots,M$, where
$$\text{cov}(Z_m,Z_{m^\prime}\,|\,\pmb{Y}_1 )={\sum_{i=1}^{k_2}}\frac{c_{mi2} c_{m^\prime i2}}{n_{i2}} \Bigg/
\sqrt{\left( {\sum_{i=1}^{k_1}}\frac{c_{mi1}^2}{n_{i1}}+ {\sum_{i=1}^{k_2}}\frac{c_{mi2}^2}{n_{i2}}\right)\left( {\sum_{i=1}^{k_1}}\frac{c_{m^\prime i1}^2}{n_{i1}}+ {\sum_{i=1}^{k_2}} \frac{c_{m^\prime i2}^2}{n_{i2}}\right)}.$$
Use of $\tilde{u}_{1-\alpha}$ as the critical value for the AMCT controls the Type I error probability at level $\alpha$ (M\"{u}ller and Sch\"{a}fer, 2001, 2004; Miller, 2010).

\subsection{Unknown Variance Case}\label{AMCT, Unknown Variance Case}

Miller (2010) briefly discusses the possibility of extending the AMCT to accommodate estimation of the variance $\sigma^2$, the complication being that the conditional Type I error probability depends on the unknown variance. Posch {\it{et al}}. (2004) developed methods to calculate the conditional Type I error probability for the one sample $t$-test given the interim data, but the authors only consider the univariate case and the approach does not directly apply to either the single contrast test or the multiple contrast test.

In this subsection, we extend the AMCT to the unknown variance case by considering the combined $T$-statistics
$$T_m=\frac{\displaystyle{\sum_{i=1}^{k_1}} c_{mi1}\bar{Y}_{i1}+\displaystyle{\sum_{i=1}^{k_2}} c_{mi2} \bar{Y}_{i2}}{S\sqrt{\displaystyle{\sum_{i=1}^{k_1}}  \frac{c^2_{mi1}}{n_{i1}}+\displaystyle{\sum_{i=1}^{k_2}}\frac{c^2_{mi2}}{n_{i2}}}}=\frac{\sigma Z_m}{S},\quad m= 1, \ldots ,M,$$
where the pooled variance estimator is
$$S^2=\left(\sum_{i=1}^{k_1}\sum_{j=1}^{n_{i1}} (Y_{ij1}-\bar{Y}_{i1})^2+\sum_{i=1}^{k_2}\sum_{j=1}^{n_{i2}} (Y_{ij2}-\bar{Y}_{i2})^2\right)\Bigg/\left(\sum_{i=1}^{k_1} n_{i1}-k_1+\sum_{i=1}^{k_2} n_{i2}-k_2\right).$$
As in Section \ref{AMCT, Known Variance Case}, we pre-specify $c^*_{mi2}=c_{mi1}$, $k^*_2=k_1$, $n^*_{i2}=n_{i1}$, and $\pmb{d}^*_{\text{Stage2}}=\pmb{d}_{\text{Stage1}}$, $i=1,\ldots,k^*_2$, $m=1,\ldots,M$. The $T$-statistics for the base test are
$$T^*_m=\sum_{i=1}^{k_1} c_{mi1} (\bar{Y}_{i1}+\bar{Y}_{i2})\Bigg/S^*\sqrt{2\sum_{i=1}^{k_1}\frac{c^2_{mi1}}{n_{i1}}}=\frac{\sigma Z^*_m }{S^*},\quad m=1,\ldots,M,$$
where
$$S^{*2}=\sum_{i=1}^{k_1}\sum_{j=1}^{n_{i1}}\left[(Y_{ij1}-\bar{Y}_{i1})^2+(Y_{ij2}-\bar{Y}_{i2})^2\right]\Bigg/(2\nu_1),\text{ where }\nu_1=\sum_{i=1}^{k_1} n_{i1}-k_1.$$
Since $S^{*2}$ is independent of $Z^*_m$ and $2\nu_1 S^{*2}/\sigma^2 \sim \chi^2_{2 \nu_1}$, the null joint distribution of $\pmb{T}^*=(T_1^*, \ldots, T_M^*)^\prime$ is multivariate $t$ with $2\nu_1$ degrees of freedom and correlation matrix $\pmb{R}^*$. The non-adaptive $\alpha$-level critical value $c^*_{1-\alpha}$ can then be obtained using the {\tt{qmvt}} function in the {\tt{R}}-package {\tt{mvtnorm}}.

The main difficulty in the unknown variance case is that the approach outlined in Section \ref{AMCT, Known Variance Case} cannot be employed because the conditional distribution of $T_m^*$ given $\pmb{Y}_1$ is not central $t$ under $H_0$. We develop the conditional Type I error probability as follows. Denote
$$T_m^*\,|\,\pmb{Y}_1=
\frac{\displaystyle{\sum_{i=1}^{k_1}} c_{mi1} (\bar{y}_{i1}+\bar{Y}_{i2})}{\sqrt{\displaystyle{\sum_{i=1}^{k_1}}\frac{c^2_{mi1}}{n_{i1}}}\sqrt{ \displaystyle{\sum_{i=1}^{k_1}\sum_{j=1}^{n_{i1}}}\left\{ (y_{ij1}-\bar{y}_{i1})^2+ (Y_{ij2}-\bar{Y}_{i2})^2\right\} \Bigg/\nu_1}}=\frac{U^*_m}{\displaystyle{\sqrt{\frac{V^*}{\nu_1}+q^*}}},\quad m=1,\ldots, M,$$
where
$$U_m^*=\frac{\displaystyle{\sum_{i=1}^{k_1}} c_{mi1} (\bar{y}_{i1}+\bar{Y}_{i2})}{\sigma\sqrt{\displaystyle{\sum_{i=1}^{k_1}}\frac{c^2_{mi1}}{n_{i1}}}},\quad V^*=\sum_{i=1}^{k_1}\sum_{j=1}^{n_{i1}} (Y_{ij2}-\bar{Y}_{i2})^2 \Big/\sigma^2,$$
and the constant
$$q^*= {\sum_{i=1}^{k_1}\sum_{j=1}^{n_{i1}}} (y_{ij1}-\bar{y}_{i1})^2\Big/(\nu_1\sigma^2).$$

Under $H_0$, the joint distribution of $(U_1^*,\ldots,U_M^*)^\prime$ is multivariate normal with mean vector $(b^*_1,\ldots, b^*_M)^\prime$ and variance-covariance matrix $\pmb{R}^*$, where
$$b^*_m=\sum_{i=1}^{k_1} c_{mi1}\bar{y}_{i1}\Bigg/\sigma\sqrt{\sum_{i=1}^{k_1}\frac{c^2_{mi1}}{n_{i1}}},\quad m=1,\ldots,M.$$
Since $V^*\sim\chi^2_{\nu_1}$ and is independent of $(U_1^*, \ldots,U_M^*)^\prime$, the joint density function of $(U_1^*,\ldots,U_M^*,V^*)^\prime$ is
\begin{eqnarray*}
& &f_{(U_1^*,\ldots,U_M^*,V^*)}(u_1^*,\ldots,u_M^*,v^*)=\frac{1}{(2 \pi)^{M/2} |\pmb{R}^*|^{1/2}}\frac{1}{\Gamma(\nu_1/2) 2^{\nu_1/2}}\times
\\[0.5\baselineskip]
& &(v^*)^{\nu_1/2-1}e^{-v^*/2}\exp\left \{-\frac{1}{2}(u^*_1-b^*_1,\ldots,u^*_M-b^*_M)(\pmb{R}^*)^{-1}(u^*_1-b^*_1,\ldots,u^*_M-b^*_M)^\prime\right \},
\end{eqnarray*}
where $\Gamma (\cdot)$ is the Gamma function. Now make the transformation
$$T_m^*\,|\,\pmb{Y}_1=\frac{U_m^*}{\displaystyle{\sqrt{\frac{V^*}{\nu_1}+q^*}}},\quad m=1,\ldots,M,\quad\text{and}\quad
W^*=V^*$$
with Jacobian $(W^*/\nu_1+q^*)^{M/2}$. The joint density function of $\pmb{T}^*\,|\,\pmb{Y}_1$ is
\begin{eqnarray*}
& &\displaystyle{f_{\pmb{T}^*\,|\,\pmb{Y}_1}\left((t_1^*,\ldots,t_M^*)\,|\,\pmb{y}_1 \right)} \\
&=&\frac{1}{(2\pi)^{M/2} |\pmb{R}^*|^{1/2}}\frac{1}{\Gamma (\nu_1/2) 2^{\nu_1/2}}
\int_0^{+ \infty}\left(\frac{w^*}{\nu_1}+q^*\right)^{M/2}(w^*)^{\nu_1/2-1} e^{-w^*/2}\times \\
& &\exp\Bigg[-\frac{1}{2}\left\{t_1^*\left(\frac{w^*}{\nu_1}+q^*\right)^{1/2}-b^*_1,\ldots,t_M^*\left(\frac{w^*}{\nu_1}+q^*\right)^{1/2}-b^*_M\right \}(\pmb{R}^*)^{-1} \\
& &\left \{t_1^*\left(\frac{w^*}{\nu_1}+q^*\right)^{1/2}-b^*_1,\ldots,t_M^*\left(\frac{w^*}{\nu_1}+q^*\right)^{1/2}-b^*_M\right\}^\prime\, \Bigg ] dw^*.%\bigg|\,\pmb{y}_1
\end{eqnarray*}
We then obtain the conditional Type I error probability
\begin{eqnarray*}
A&=&1- P_{H_0}\left(\pmb{T}^*\leq (c^*_{1-\alpha},\ldots,c^*_{1-\alpha})^\prime\,|\,\pmb{Y}_1\right) \\
&=&1- \int\cdots\int_{(t_1^*,\ldots, t_M^*) \leq (c^*_{1-\alpha},\ldots,c^*_{1-\alpha})} f_{\pmb{T}^*\,|\,\pmb{Y}_1}\left((t_1^*,\ldots,t_M^*)\,|\,\pmb{y}_1\right)\ d t_1^* \cdots d t_M^*.
\end{eqnarray*}

After making the adaptations at the interim analysis, from the conditional distribution of $\pmb{T}=(T_1, \ldots, T_M)^\prime$ given $\pmb{Y}_1$, the adaptive critical value $\tilde{c}_{1-\alpha}$ can be determined as a solution to the following equation:
\begin{eqnarray*}
\tilde{A}&=&1-P_{H_0}\left(\pmb{T} \leq (\tilde{c}_{1-\alpha},\ldots,\tilde{c}_{1-\alpha})^\prime\,|\,\pmb{Y}_1\right) \\
&=&1- \int\cdots\int_{(t_1,\ldots,t_M) \leq (\tilde{c}_{1-\alpha},\ldots,\tilde{c}_{1-\alpha})} f_{\pmb{T}\,|\,\pmb{Y}_1}\left((t_1,\ldots,t_M)\,|\,\pmb{y}_1\right)\ d t_1 \cdots d t_M=A,
\end{eqnarray*}
where
\begin{eqnarray*}
&\displaystyle f_{\pmb{T}\,|\,\pmb{Y}_1}\left((t_1, \ldots, t_M)\,|\,\pmb{y}_1\right)=\displaystyle\frac{1}{(2\pi)^{M/2} |\pmb{\tilde{R}}|^{1/2}}\displaystyle \frac{1}{\Gamma (\nu_2/2) 2^{\nu_2/2}}\int_0^{+ \infty}\left(\frac{w}{\nu}+q\right)^{M/2} w^{\nu_2/2-1} e^{-w/2}\times \\
&\exp\Bigg[-\displaystyle\frac{1}{2}\left\{t_1\left(\frac{w}{\nu}+q\right)^{1/2}-b_1,\ldots,t_M\left(\frac{w}{\nu}+q\right)^{1/2}-b_M\right\}\pmb{\tilde{R}}^{-1} \\
&\displaystyle\left\{t_1\left(\frac{w}{\nu}+q\right)^{1/2}-b_1,\ldots,t_M\left(\frac{w}{\nu}+q\right)^{1/2}-b_M\right\}^\prime\, \Bigg] dw,\\
&\nu_2=\displaystyle{\sum_{i=1}^{k_2}} n_{i2}-k_2, \quad \nu=\nu_1+\nu_2, \quad q=\displaystyle{\sum_{i=1}^{k_1}\sum_{j=1}^{n_{i1}}} (y_{ij1}-\bar{y}_{i1})^2\Big/(\nu\sigma^2),
\end{eqnarray*}
and
\[b_m=\displaystyle{\sum_{i=1}^{k_1}} c_{mi1}\bar{y}_{i1}\Bigg/\sigma\sqrt{\displaystyle{\sum_{i=1}^{k_1}}\frac{c^2_{mi1}}{n_{i1}}+\displaystyle{\sum_{i=1}^{k_2}}\frac{c^2_{mi2}}{n_{i2}}},\quad m=1,\ldots,M.\]
$H_0$ is rejected if $T_{\max}=\max \{\pmb{T}\}\geq\tilde{c}_{1-\alpha}$. Use of the critical value $\tilde{c}_{1-\alpha}$ provides control of the Type I error probability at level $\alpha$ according to the CRP principle (M\"{u}ller and Sch\"{a}fer, 2001, 2004).

\section{Numerical Example}\label{Numerical Example}

\subsection{Adaptive Generalized Multiple Contrast Tests}\label{Numerical Example, AGMCT}

To illustrate the adaptive generalized multiple contrast tests (AGMCTs), we generated a numerical example. The example data set is available as Supporting Information. Suppose that there are $k_1=5$ dosage groups in Stage 1, with $\pmb{d}_{\text{Stage1}}=(0,0.05,0.20,0.60,1.00)^\prime$. The total sample sizes in two stages are the same ($N_1=N_2=120$) and the group sample sizes are equal in Stage 1 ($n_{11}=\cdots =n_{51}=N_1/5=24$). The $M=5$ candidate dose-response models with the original specifications of $\pmb{\theta}$ are shown in Table \ref{Table.original candidate models}.

We assume that the true dose-response model is the $E_{\max}$ 2 model:
$$f_{E_{\max}2}(d,\pmb{\theta})=E_0+E_{\text{max}} d/ (ED_{50}+d)=0.2+0.6 d/(0.1+d).$$
We generate the Stage 1 data from a multivariate normal distribution with mean $f_{E_{\max}2}(\pmb{d}_{\text{Stage1}},\pmb{\theta})=$ $(0.20, 0.40, 0.60, 0.71, 0.75)^\prime$ and covariance matrix $\sigma^2\pmb{I}=1.478^2\pmb{I}$. The sample mean and variance estimates from the Stage 1 data are $\bar{\pmb{y}}_1=(0.52, 0.47, 1.09, 1.70, 0.45)^\prime$ and $s^2_1=1.58^2$, respectively.

The optimal contrast vectors in Stage 1 based on the $M=5$ candidate dose-response models in Table \ref{Table.original candidate models} are as follows.
\begin{align*}
&E_{\max}: \pmb{c}_{11}=(-0.64, -0.36, 0.06, 0.41, 0.53)^\prime, \\
&\text{Linear-log}: \pmb{c}_{21}=(-0.54, -0.39, -0.08, 0.37, 0.64)^\prime, \\
&\text{Linear}: \pmb{c}_{31}=(-0.44, -0.38, -0.20, 0.27, 0.74)^\prime, \\
&\text{Quadratic}: \pmb{c}_{41}=(-0.57, -0.36,  0.16, 0.71, 0.07)^\prime, \\
&\text{Logistic}: \pmb{c}_{51}=(-0.40, -0.39, -0.31, 0.50, 0.59)^\prime.\end{align*}
After conducting three different GMCTs using Tippett's, Fisher's, and inverse normal combination statistics, we obtain the following Stage 1 p-values: $p_{T1}=0.005$, $p_{F1}=0.047$, and $p_{N1}=0.06$.

%%%%%%%%%%%%%%%%%%%%%%%%%%%%%%%%%%%%%%%%%%%%%%%%%%%%%%%%%%%%%%%%%%%%%
We then adapt the candidate dose-response models and the dosage groups. We fit the 5 original candidate dose-response models using the Stage 1 data. Unfortunately, the Logistic model failed to converge on a solution so we replaced it with isotonic regression. Also, we use the dosage adaptation rule described in Section \ref{Adapting the Dosage Groups} with $\delta=0$ to drop the active dosage groups that appear to be less efficacious than placebo or the adjacent dosage. Finally, we obtain $k_2=3$ dosage groups in Stage 2: $\pmb{d}_{\text{Stage2}}=(0,0.20,0.60)^\prime$ and $n_{12}=n_{22}=n_{32}=N_2/3=40$.

The optimal contrast vectors in Stage 2 based on the adapted dose-response models and dosage groups are as follows:
\begin{align*}
&E_{\max}: \pmb{c}_{12}=(-0.433, -0.383, 0.816)^\prime,\\
&\text{Linear-log}: \pmb{c}_{22}=(-0.707, 0.000, 0.707)^\prime,\\
&\text{Linear}: \pmb{c}_{32}=(-0.617, -0.154, 0.772)^\prime, \\
&\text{Quadratic}: \pmb{c}_{42}=(-0.766, 0.137, 0.629)^\prime,\\
&\text{Isotonic regression}: \pmb{c}_{52}=(-0.816, 0.408, 0.408)^\prime.\end{align*}

The Stage 2 data are then generated from a multivariate normal distribution with mean $f_{E_{\max}2}(\pmb{d}_{\text{Stage2}},\pmb{\theta})=$ $(0.20, 0.60, 0.71)^\prime$ and covariance matrix $\sigma^2\pmb{I}=1.478^2\pmb{I}$. The sample mean and variance estimates from the Stage 2 data under adaptation are $\bar{\pmb{y}}_2=(-0.09, 0.77, 0.73)^\prime$ and $s^2_2=1.52^2$, respectively. After conducting three different GMCTs using Tippett's, Fisher's, and inverse normal combination statistics, we obtain the following Stage 2 p-values: $p_{T2}=0.005$, $p_{F2}=0.008$, and $p_{N2}=0.008$. The p-values from Stage 1 and Stage 2 are then combined using Fisher's combination statistic and the inverse normal combination statistic. The combination statistics and resulting overall p-values are shown in Table \ref{num.results AGMCT}.

\subsection{Adaptive Multiple Contrast Test}\label{Numerical Example, AMCT}

\subsubsection{Known Variance Case}
We use the same simulated data as in Section \ref{Numerical Example, AGMCT} to illustrate the adaptive multiple contrast test (AMCT) for the known variance case (for purposes of this illustration, we use $\sigma^2=1.478^2$). We first obtain the non-adaptive critical value $u^*_{1-\alpha}$. The joint null distribution of $\pmb{Z}^*=(Z_1^*,\ldots,Z_5^*)^\prime$ is multivariate normal with mean $\pmb{0}$ and covariance matrix $\pmb{R}^*$, where
\begin{eqnarray*}
\pmb{R}^*=\left(\begin{array}{ccccc}
1 &0.977 &0.912& 0.842& 0.896 \\
0.977 &1& 0.977& 0.750& 0.956 \\
0.912 &0.977& 1& 0.602& 0.957 \\
0.842 &0.750& 0.602& 1& 0.715 \\
0.896 &0.956& 0.957& 0.715& 1
\end{array}\right).
\end{eqnarray*}

The value of $u^*_{1-\alpha}$ is obtained using the {\tt{qmvnorm}} function in the {\tt{R}}-package {\tt{mvtnorm}}, resulting in $u^*_{1-\alpha}=1.968$. We then calculate the conditional mean of $\pmb{Z}^*$ given $\pmb{Y}_1$,
$$\left(\frac{\displaystyle{\sum_{i=1}^{k_1}} c_{1i1} \bar{y}_{i1}}{\sigma\sqrt{2\displaystyle{\sum_{i=1}^{k_1}}\frac{c_{1i1}^2}{n_{i1}}}},\ldots,
\frac{\displaystyle{\sum_{i=1}^{k_1}} c_{Mi1} \bar{y}_{i1}}{\sigma\sqrt{2\displaystyle{\sum_{i=1}^{k_1}}\frac{c_{Mi1}^2}{n_{i1}}}}\right)^\prime=(1.19, 0.87, 0.42, 2.22, 0.92)^\prime,$$
and the conditional covariance matrix $\pmb{R_2}^*=\pmb{R}^*/2$. The conditional error is obtained using the {\tt{pmvnorm}} function in the {\tt{R}}-package {\tt{mvtnorm}} as
$$A=1- P_{H_0}\left(\pmb{Z}^*  \leq (u^*_{1-\alpha},\ldots, u^*_{1-\alpha})^{\prime}\,|\,\pmb{Y}_1\right)=0.64.$$

After adapting the dose-response models and dosage groups as in Section \ref{Numerical Example, AGMCT} above, we obtain the conditional distribution of $\pmb{Z}\,|\,\pmb{Y}_1$, which is multivariate normal with mean
$$
\left(\frac{\displaystyle{\sum_{i=1}^{k_1}} c_{1i1} \bar{y}_{i1}}{\sigma\sqrt{\displaystyle{\sum_{i=1}^{k_1}}\frac{c_{1i1}^2}{n_{i1}}+\displaystyle{\sum_{i=1}^{k_2}}\frac{c_{1i2}^2}{n_{i2}}}},\ldots,\frac{\displaystyle{\sum_{i=1}^{k_1}} c_{Mi1} \bar{y}_{i1}}{\sigma\sqrt{\displaystyle{\sum_{i=1}^{k_1}}\frac{c_{Mi1}^2}{n_{i1}}+\displaystyle{\sum_{i=1}^{k_2}} \frac{c_{Mi2}^2}{n_{i2}}}},\right)^{\prime} \\[0.5\baselineskip]=(1.33, 0.98, 0.47, 2.48, 1.03)^{\prime}
$$
and covariance matrix
\begin{eqnarray*}
\pmb{\tilde{R}}=\left(\begin{array}{ccccc}
0.375& 0.331& 0.358& 0.297& 0.199 \\
0.331& 0.375& 0.368& 0.370& 0.325 \\
0.358& 0.368& 0.375& 0.351& 0.283 \\
0.297& 0.370& 0.351& 0.375& 0.352 \\
0.199& 0.325& 0.283& 0.352& 0.375
\end{array}\right).
\end{eqnarray*}

Finally, we obtain the adaptive critical value $\tilde{u}_{1-\alpha}=2.263$ and the combined test statistics $\pmb{Z}=(Z_1,\ldots, Z_M)^\prime=(2.22, 2.50, 1.78, 4.15, 2.83)^\prime$. We reject $H_0$ since $Z_{\max}=4.15 \geq \tilde{u}_{1-\alpha}$.

\subsubsection{Unknown Variance Case}

To illustrate the AMCT in the unknown variance case (Section \ref{AMCT, Unknown Variance Case}), we use the same example data as in Section \ref{Numerical Example, AGMCT} for $M=2$ candidate dose-response models. Here, we only consider the $E_{\max}$ and Linear-log candidate dose-response models in Table \ref{Table.original candidate models}. Other settings are the same as in Section \ref{Numerical Example, AGMCT}, including the optimal contrasts for
both Stage 1 and Stage 2, and the adapted dosage groups for Stage 2.

We first obtain the non-adaptive critical value $c^*_{1-\alpha}$. The joint null distribution of $\pmb{T}^*=(T_1^*,T_2^*)^\prime$ is bivariate $t$ with degrees of freedom $2 \nu_1$ and correlation matrix $\pmb{R}^*$, where $\nu_1=N_1-5=115$ and
\begin{eqnarray*}
\pmb{R}^*=\left(\begin{array}{cc}
1 &0.977\\
0.977 &1
\end{array}\right).
\end{eqnarray*}
The value of $c^*_{1-\alpha}$ is obtained using the {\tt{qmvt}} function in the {\tt{R}}-package {\tt{mvtnorm}}, resulting in $c^*_{1-\alpha}=1.732$.

We then obtain the conditional error by numerically calculating the three-dimensional integral below using the {\tt{adaptIntegrate}} function in the {\tt{R}}-package {\tt{cubature}}.
\begin{eqnarray*}
A&=&1- \frac{1}{(2\pi)^{M/2} |\pmb{R}^*|^{1/2}}\frac{1}{\Gamma (\nu_1/2) 2^{\nu_1/2}} \int_0^{+ \infty} \int_{-\infty}^{c^*_{1-\alpha}} \int_{-\infty}^{c^*_{1-\alpha}} \left(\frac{w^*}{\nu_1}+q^*\right)^{M/2}(w^*)^{\nu_1/2-1} e^{-w^*/2}\times\\
& &\exp\Bigg[-\frac{1}{2}\left\{t_1^*\left(\frac{w^*}{\nu_1}+q^*\right)^{1/2}-b^*_1, t_2^*\left(\frac{w^*}{\nu_1}+q^*\right)^{1/2}-b^*_2\right\}(\pmb{R}^*)^{-1} \\
& & \left\{t_1^*\left(\frac{w^*}{\nu_1}+q^*\right)^{1/2}-b^*_1, t_2^*\left(\frac{w^*}{\nu_1}+q^*\right)^{1/2}-b^*_2\right\}^\prime\, \Bigg] dw^* \ dt_1^* \ dt_2^* =  0.198.
\end{eqnarray*}

After adapting the dose-response models and dosage groups at the end of Stage 1, we consider the conditional distribution of $\pmb{T}\,|\,\pmb{Y}_1$. The adaptive critical value $\tilde{c}_{1-\alpha}$ can be obtained by solving the following equation using a bisection algorithm:

\begin{eqnarray*}
\tilde{A}&=& \frac{1}{(2\pi)^{M/2} |\pmb{\tilde{R}}|^{1/2}}\frac{1}{\Gamma (\nu_2/2) 2^{\nu_2/2}} \int_0^{+ \infty} \int_{-\infty}^{\tilde{c}_{1-\alpha}} \int_{-\infty}^{\tilde{c}_{1-\alpha}} \left(\frac{w}{\nu}+q\right)^{M/2} w^{\nu_2/2-1} e^{-w/2}\times \\
& &\exp\Bigg[-\frac{1}{2}\left\{t_1\left(\frac{w}{\nu}+q\right)^{1/2}-b_1, t_2\left(\frac{w}{\nu}+q\right)^{1/2}-b_2\right\}\pmb{\tilde{R}}^{-1} \\
& & \left\{t_1\left(\frac{w}{\nu}+q\right)^{1/2}-b_1, t_2\left(\frac{w}{\nu}+q\right)^{1/2}-b_2\right\}^\prime\, \Bigg] dw \ dt_1\ dt_2 = A ,
\end{eqnarray*}
where the covariance matrix $\pmb{\tilde{R}}$ is
\begin{eqnarray*}
\pmb{\tilde{R}}=\left(\begin{array}{cc}
0.375 &0.331\\
0.331 &0.375
\end{array}\right).
\end{eqnarray*}

Finally, we obtain the adaptive critical value $\tilde{c}_{1-\alpha}=1.802$ with tolerance $10^{-7}$. The combined test statistics are $\pmb{T}=(T_1,T_2)^\prime=(2.11, 2.38)^\prime$ and we reject $H_0$ since $T_{\max}=2.38 \geq \tilde{c}_{1-\alpha}$.

\section{Simulation studies}\label{Simulation studies}

%{\color{red}\textbf{(RERUN ALL THE SIMULATION STUDIES BASED ON REVIEWER'S COMMENTS. NEW RESULTS ARE UPDATED)}}

In this section, we conduct simulation studies to compare the operating characteristics of the AGMCTs with those of the AMCT in the setting of a design that adapts both the candidate dose-response models and the dosage groups based on data from Stage 1. We also compare these with the operating characteristics of the corresponding tests in a non-adaptive design.

\begin{figure}[!h]
\begin{center}
\includegraphics[width=16cm,height=8cm]{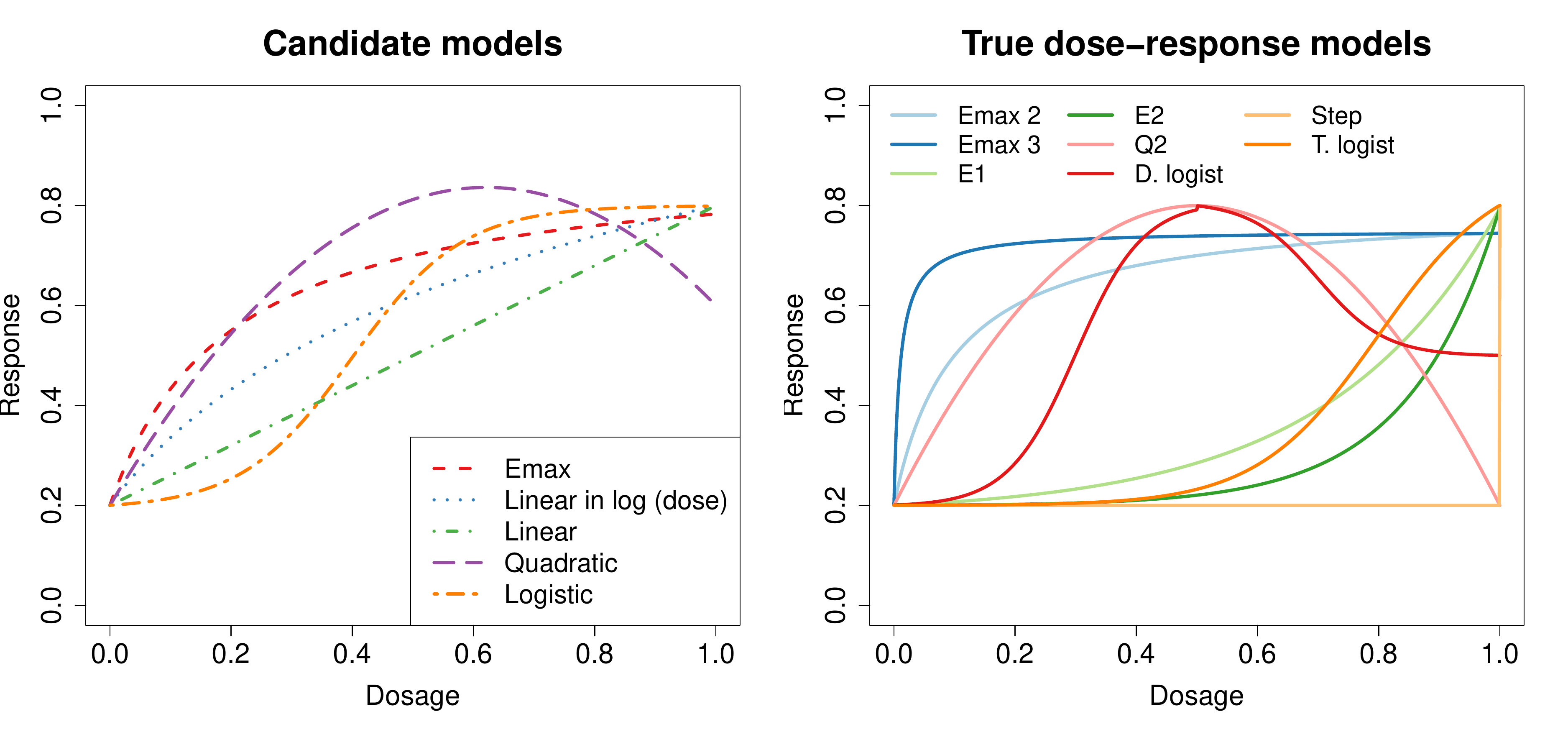}
\caption{Five candidate dose-response models (left panel) and eight true dose-response models (right panel).}\label{fig: candidate and true model}
\end{center}
\end{figure}

Assume $k_1=5$ and $\pmb{d}_{\text{Stage1}}=(0,0.05,0.20,0.60,1.00)^\prime$. The total sample size is the same for each of the two stages ($N_1=N_2$) and the group sample sizes within each stage are equal, with $N_1=N_2=60$, 120, 180, and 240. The $M=5$ candidate dose-response models with the original specifications of $\pmb{\theta}$ are shown in Table \ref{Table.original candidate models}. The outcome for each patient is distributed as $N(\mu(d),\sigma^2)$, where the true mean configuration $\mu(d)$ follows one of the eight different dose-response models in Table \ref{Sim.eight dose-response models}, and $\sigma=1.478$. The dose-response curves for the five candidate models and the eight true dose-response models are shown in Figure \ref{fig: candidate and true model}.

\begin{figure}[!h]
\begin{center}
\includegraphics[width=16cm,height=8cm]{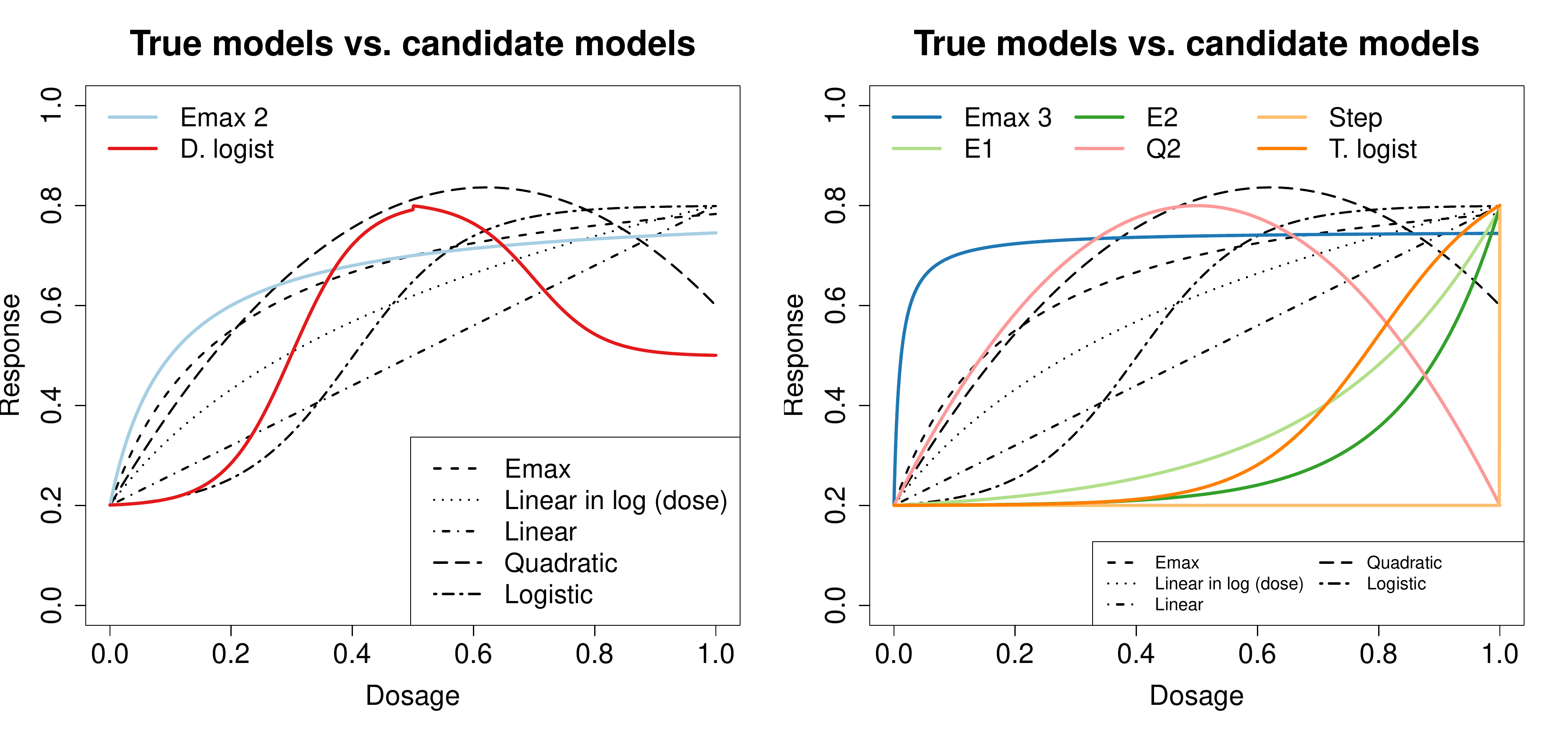}
\caption{True dose-response models vs. five candidate models. In the left panel, the optimal contrasts associated with the true dose-response models (colored) are highly correlated with those of the candidate models (black). In the right panel, the optimal contrasts associated with the true dose-response models (colored) are not highly correlated with those of the candidate models (black).}\label{fig: True vs. candidate}
\end{center}
\end{figure}

For the (true) $E_{\max}$ 2 and Double-logistic models, the optimal contrasts are highly correlated with those of the candidate models. In contrast, for the (true) $E_{\max}$ 3, Exponential 1, Exponential 2, Quadratic 2, Step and Truncated-logistic models, the optimal contrasts are not highly correlated with those of the candidate models (Figure \ref{fig: True vs. candidate}).

For the AGMCTs, we use three GMCTs to combine the $M=5$ dependent $p$-values \emph{within} each stage: Tippett's ($T$), Fisher's ($F$) and inverse normal ($N$) combination methods (Ma and McDermott, 2020). The same GMCT is used in both Stage 1 and Stage 2. To perform the overall test, only the inverse normal ($\Psi_N$) combination statistic is used to combine $p_1$ and $p_2$ \emph{across} stages since our preliminary simulation studies showed that, in general, using $\Psi_N$ to combine $p_1$ and $p_2$ yielded greater power than using $\Psi_F$. The reason for this is that under the alternative hypothesis, $p_1$ and $p_2$ both tend to be small and the rejection region of $\Psi_N$ is larger than that of $\Psi_F$ when $p_1$ and $p_2$ are both small (Wassmer and Brannath 2016, Section 6.2). %This is the circumstance under which a test based on $\Psi_N$ tends to be more powerful than a test based on $\Psi_F$. Because of this, only the results using the inverse normal combination method to combine $P$-values \emph{across} stages are presented.

%Fisher's ($\Psi_F$) and inverse normal ($\Psi_N$) combination statistics are used to combine $p_1$ and $p_2$ \emph{across} stages. For Fisher's combination statistic, we do not allow early stopping for futility and we reject the null hypothesis if $p_1 p_2 \leq \exp [\chi^2_{4,1-0.05}]=0.0087$ (Bauer and K\"{o}hne, 1994).

%The reason for this is that under the alternative hypothesis, $p_1$ and $p_2$ both tend to be small. The rejection region of $\Psi_N$ is larger than the rejection region of $\Psi_F$ when $p_1$ and $p_2$ are both small (Wassmer and Brannath 2016, Section 6.2). This is the circumstance under which a test based on $\Psi_N$ tends to be more powerful than a test based on $\Psi_F$. Because of this, only the results using the inverse normal combination method to combine $P$-values \emph{across} stages are presented.

For the AGMCTs, we report the results of the operating characteristics for both the known and unknown variance cases. The results for the corresponding GMCTs in a non-adaptive design are also reported. For the AMCT, the simulation studies of the operating characteristics are presented only for the known variance case. The corresponding test in a non-adaptive design is just the MCP-Mod procedure, which is equivalent to the GMCT based on Tippett's combination method in a non-adaptive design.

All dosage adaptations are made according to the example rule described in Section \ref{Adapting the Dosage Groups}. To deal with the problems outlined in Section \ref{Adapting the Candidate Dose-Response Models} above, if only one of the $E_{\max}$ and Logistic models fails to converge in Stage 1, isotonic regression is used to generate the corresponding contrast for use in Stage 2; if both the $E_{\max}$ and Logistic models fail to converge in Stage 1, then isotonic regression is used to generate the corresponding contrast for the Logistic model and the same contrast that was used in Stage 1 is used in Stage 2 for the $E_{\max}$ model.  Also, if there is a negative dose-response relationship suggested by the Stage 1 data (i.e., a negative estimated slope in the Linear model), no adaptation of the dose-response models is performed for Stage 2 and we only adapt the dosage groups.

All estimated values of Type I error probability and power are based on 10,000 replications of the simulations. The Type I error probabilities for the AGMCTs and the AMCT (Tables \ref{sim: type I error, known variance} and \ref{sim: type I error, unknown variance} in the Appendix) agree with theory that the tests being considered all exhibit control of the Type I error probability at $\alpha=0.05$; all values fall within the 95\% confidence interval (0.0457, 0.0543).

For the known variance case, the power curves of the competing tests are shown in Figure \ref{fig: Power AGMCTs and AMCT, known variance}. When the optimal contrasts associated with the true dose-response models are highly correlated with those of the candidate models ($E_{\max}$ 2 and Double-logistic models), the AGMCTs and the AMCT are, in general, slightly less powerful than the corresponding tests in a non-adaptive design. When the optimal contrasts associated with the true dose-response models are not highly correlated with those of the candidate models ($E_{\max}$ 3, Exponential 1, Exponential 2, Quadratic 2, Step and Truncated-logistic models), however, the AGMCTs and AMCT are more powerful than the corresponding tests in a non-adaptive design. Another observation is that the overall performance of the AMCT is the best among all the adaptive designs. %to those of the AGMCTs

For the unknown variance case, the power curves of the competing tests are shown in Figure \ref{fig: Power AGMCTs, unknown variance}. The overall results for these comparisons are very similar to those for the known variance case.

\section{Conclusion}\label{Conclusion}

In this article, we extend the MCP-Mod procedure with GMCTs (Bretz {\it{et al}}., 2005; Ma and McDermott, 2020) to two-stage adaptive designs. We perform a GMCT within each stage and combine the stage-wise $p$-values using a specified combination method to test the overall null hypothesis of no dose-response relationship. We also consider and extend an alternative AMCT approach proposed by Miller (2010), which uses the maximum standardized stratified contrast across Stage 1 and Stage 2 as the test statistic. One issue that deserves further exploration is how to best determine the ``base test" for the AMCT. Our development in Sections \ref{AMCT, Known Variance Case} and \ref{AMCT, Unknown Variance Case} is based on pre-specification of the contrasts, number of candidate dose-response models, and group sample sizes to be the same in Stage 2 as they were in Stage 1. While this is not necessarily the best choice, in the absence of the ability to enumerate all possible two-stage designs being considered, it might be quite reasonable in practice. An issue that remains unresolved is that of efficiently computing the conditional error and adaptive critical value for the AMCT when the variance is unknown since these involve multidimensional integrals that can take a long time to compute.

Simulation studies demonstrate that the AGMCTs and AMCT are generally more powerful for PoC testing than the corresponding tests in a non-adaptive design if the true dose-response model is, in a sense, not ``close'' to the models included in the initial candidate set. This might occur, for example, if the selection of the candidate set of dose-response models is not well informed by evidence from preclinical and early-phase studies. This is consistent with intuition: if the dose-response models are badly misspecified at the design stage, using data from Stage 1 to get a better sense of the true dose-response model and using data from both Stage 1 and Stage 2 to perform an overall test for $H_0$ should result in increased power. On the other hand, if the true dose-response model is ``close" to the models specified in the initial candidate set, the non-adaptive design is sufficient to detect the PoC signal. In this case, the adaptive design does not provide any benefit and results in a small loss of efficiency.

Comparisons among the different AGMCTs and the AMCT did not reveal major differences in their operating characteristics in general. Differences among the AGMCTs tended to be larger in the setting of a non-adaptive design (Ma and McDermott, 2020). In principle, the AGMCTs proposed here for two-stage adaptive designs could be extended to multiple stages, although the circumstances under which that would be beneficial are not clear.

Finally, we note that baseline covariates can easily be incorporated into the AGMCTs, as outlined in Section 2.3 of Ma and McDermott (2020).\\

\setstretch{1}

\makeatletter
\renewcommand{\@biblabel}[1]{#1.}

\bigskip

\noindent {\bf{Conflict of Interest}}

\noindent {\it{The authors have declared no conflict of interest. }}

\newpage

\newpage

\begin{figure}[!h]
\begin{center}
\includegraphics[width=12cm,height=21cm]{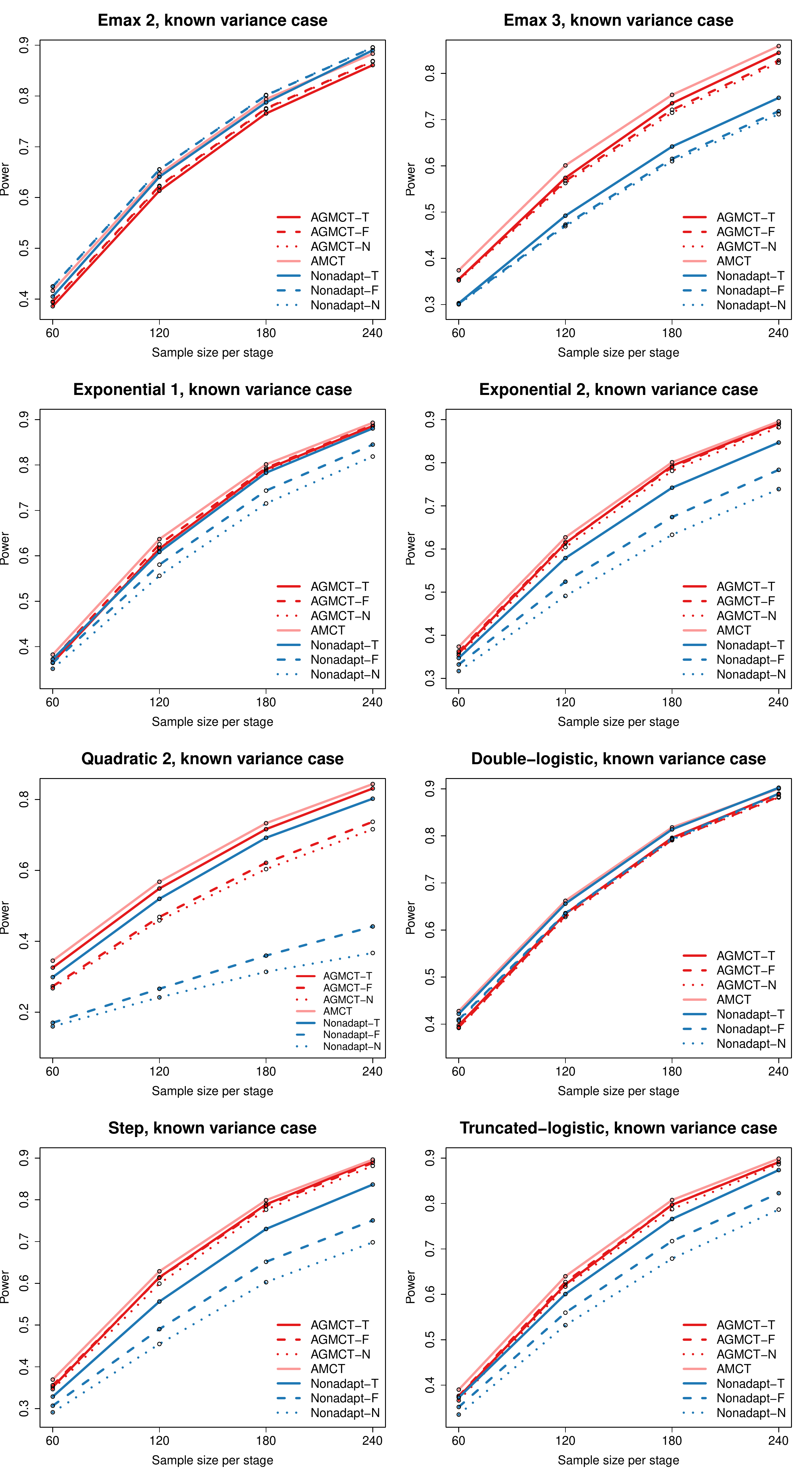}
\caption{Power curves for the AGMCTs and the AMCT in the known variance case for designs that adapt the candidate dose-response models, as well as the corresponding tests in a non-adaptive design.}\label{fig: Power AGMCTs and AMCT, known variance}
\end{center}
\end{figure}

\newpage

\begin{figure}[!h]
\begin{center}
\includegraphics[width=12cm,height=21cm]{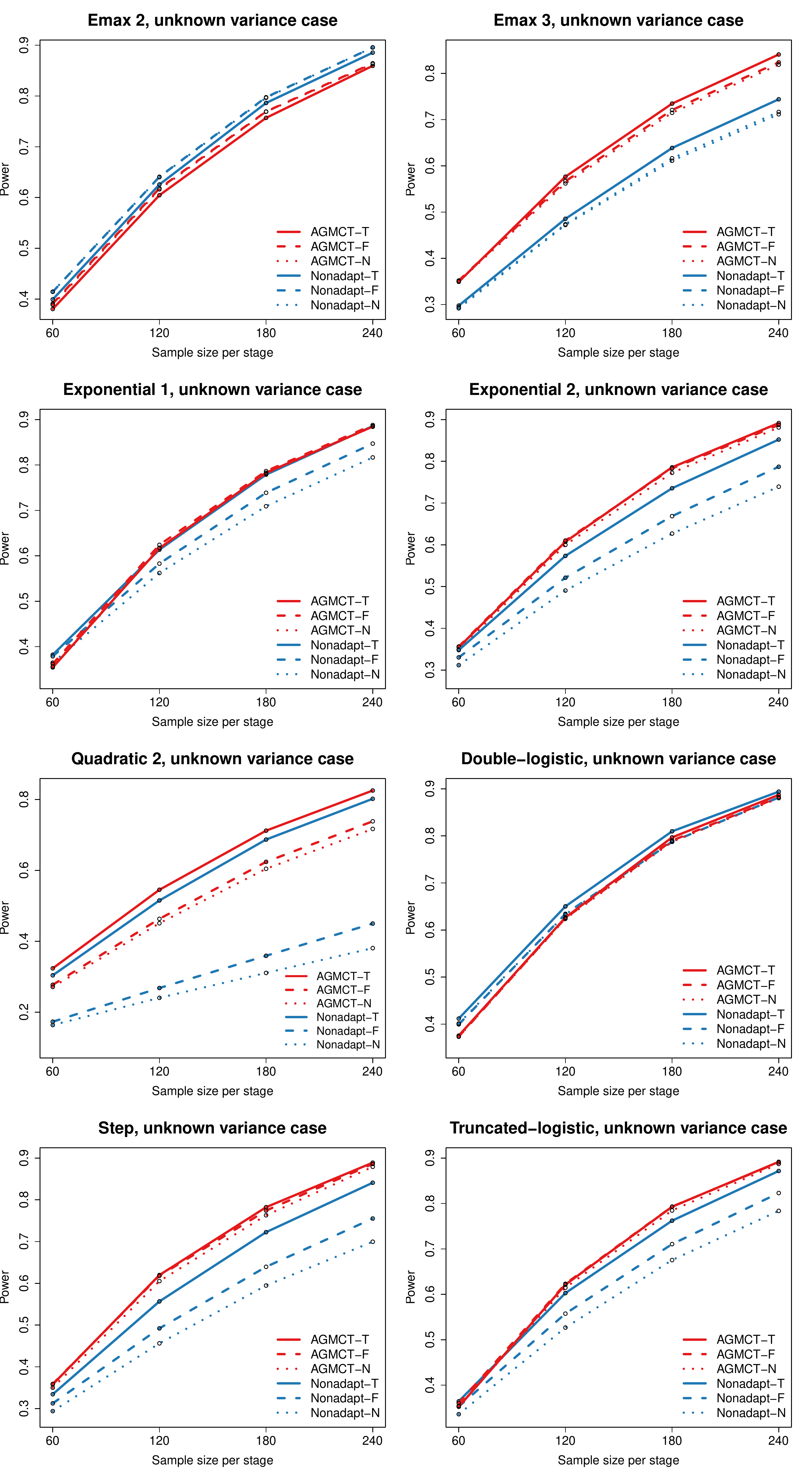}
\caption{Power curves for the AGMCTs in the unknown variance case for designs that adapt the candidate dose-response models, as well as the corresponding tests in a non-adaptive design.}\label{fig: Power AGMCTs, unknown variance}
\end{center}
\end{figure}

\newpage
\begin{table}[!htbp]
\footnotesize
\begin{center}
\caption{$M=5$ original candidate dose-response models.}\label{Table.original candidate models}
\begin{tabular}{ll}
\hline
$E_{\max}$  & $f_1(d,\pmb{\theta})=E_0+E_{\text{max}} d/ (ED_{50}+d)= 0.2+0.7 d/(0.2+d)$   \\ \hline
Linear-log & $f_2 (d,\pmb{\theta})=\theta_0+\theta_1 \log(5d+1)=0.2+ \{0.6/\log(6)\} \log(5d+1) $ \\ \hline
Linear & $f_3 (d,\pmb{\theta})=\theta_0+\theta_1 d=0.2+0.6 d$ \\ \hline
Quadratic& $f_4 (d,\pmb{\theta})=\theta_0+\theta_1 d + \theta_2 d^2=0.2+ 2.049 d-1.749 d^2$ \\ \hline
Logistic & \begin{tabular}[c]{@{}l@{}}
$f_5 (d,\pmb{\theta})=E_0+E_{\text{max}} /[1 + \exp\{(ED_{50}-d)/ \delta\} ]$ \\
  $= 0.193+0.607/[1+\exp\{(0.4-d)/ 0.09\} ]$ \end{tabular} \\ \hline
\end{tabular}
\end{center}
\end{table}

\begin{table}[!htbp]
\begin{center}
\caption{Combining $p_1$ and $p_2$ across stages using Fisher's and inverse normal combination methods.}\label{num.results AGMCT}
\begin{tabular}{cccc|cccc}
\hline
\multicolumn{4}{c|}{Fisher} & \multicolumn{4}{c}{Inverse Normal} \\ \hline
\begin{tabular}[c]{@{}c@{}} Within-stage\\ combination \\statistic\end{tabular}  &\begin{tabular}[c]{@{}c@{}} Across \\stages \\ $\Psi_F$ \end{tabular}  & \begin{tabular}[c]{@{}c@{}} Overall\\ p-value\end{tabular} & Reject $H_0$&
\begin{tabular}[c]{@{}c@{}} Within-stage\\ combination \\statistic\end{tabular}  & \begin{tabular}[c]{@{}c@{}} Across \\stages \\ $\Psi_N$ \end{tabular}  & \begin{tabular}[c]{@{}c@{}} Overall\\ p-value\end{tabular} & Reject $H_0$\\ \hline
$\Psi_T$ & 21.23 & 0.0003 &Yes &$\Psi_T$ & 5.16 & 0.0001 &Yes\\ \hline
$\Psi_F$ & 15.78 & 0.003 &Yes &$\Psi_F$ & 4.08 & 0.002 &Yes\\ \hline
$\Psi_N$ & 15.18 & 0.004 &Yes &$\Psi_N$ & 3.95 & 0.003 &Yes\\ \hline
\end{tabular}
\end{center}
\end{table}

\begin{table}[!htbp]
\footnotesize
\begin{center}
  \caption{Eight different true dose-response models considered in the simulation studies.}\label{Sim.eight dose-response models}
\begin{tabular}{ll}
\hline
$E_{\max}$ 2  & $0.2+0.6 d/(0.1+d)$ \\ \hline
$E_{\max}$ 3  & $0.2+0.55 d/(0.01+d)$ \\ \hline
Exponential 1 & $0.183+0.017 \ \exp\{2d\ \log(6)\}$ \\ \hline
Exponential 2 & $0.19924+0.00076 \ \exp(d/0.15)$ \\ \hline
Quadratic 2 & $0.2+2.4 d- 2.4 d^2$ \\ \hline
Double-logistic & \begin{tabular}[c]{@{}l@{}}
$\left[0.198+\displaystyle{\frac{0.61}{1+\exp\{18 (0.3-d)\}}}\right\} I(d\leq 0.5)$ \\
  $+\left\{0.499+\displaystyle{\frac{0.309}{1+\exp\{18 (d-0.7)\}}}\right] I(d>0.5)$ \end{tabular} \\ \hline
Step & $0.2+0.6 I(d \geq 0.6)$ \\ \hline
Truncated-logistic & $0.2+0.682/\left[1+\exp\{10 (0.8-d)\}\right]$ \\ \hline
\end{tabular}
\end{center}
\end{table}

\label{lastpage}

\clearpage
\renewcommand{\thepage}{A\arabic{page}}
\renewcommand{\thesection}{A\arabic{section}}
\renewcommand{\thetable}{A\arabic{table}}
\renewcommand{\thefigure}{A\arabic{figure}}

\setcounter{page}{1}
\setcounter{section}{1}
\setcounter{figure}{0}
\setcounter{table}{0}

\maketitle
\section*{Appendix} % {\it(please insert here, if applicable)}

In this section, we display the Type I error probabilities of the AGMCTs and the AMCT for the known and unknown variance cases in Tables \ref{sim: type I error, known variance} and \ref{sim: type I error, unknown variance}, respectively.

\begin{table}[!htbp]
\scriptsize
\begin{center}
\caption{Type I error probabilities of the AGMCTs and the AMCT in the known variance case for designs that adapt the candidate dose-response models, as well as the corresponding tests in a non-adaptive design.}\label{sim: type I error, known variance}
\begin{tabular}{|l|l|l|l|l|}
\hline
\multicolumn{5}{|c|}{$N_1=N_2=60$}   \\ \hline
& \multicolumn{3}{c|}{AGMCT} & AMCT \\ \hline
& T      & F     & N     &      \\ \hline
Adaptive&0.0487 &0.0519 &0.0515 &0.0510   \\ \hline
Non-adaptive &0.0530 &0.0523& 0.0520 &0.0533 \\ \hline
\multicolumn{5}{|c|}{$N_1=N_2=120$}   \\ \hline
& \multicolumn{3}{l|}{AGMCT} & AMCT \\ \hline
& T      & F      & N    &      \\ \hline
Adaptive&0.0470 &0.0468& 0.0467& 0.0463\\ \hline
Non-adaptive &0.0479 &0.0487 &0.0485 &0.0480 \\ \hline
\multicolumn{5}{|c|}{$N_1=N_2=180$}   \\ \hline
& \multicolumn{3}{c|}{AGMCT} & AMCT \\ \hline
& T     & F      & N    &      \\ \hline
Adaptive& 0.0507& 0.0486 &0.0488 &0.0515  \\ \hline
Non-adaptive & 0.0492& 0.0470& 0.0472 &0.0491\\ \hline
\multicolumn{5}{|c|}{$N_1=N_2=240$}   \\ \hline
& \multicolumn{3}{c|}{AGMCT} & AMCT \\ \hline
& T      & F     & N     &      \\ \hline
Adaptive & 0.0507 &0.0502& 0.0492& 0.0503   \\ \hline
Non-adaptive &0.0525 &0.0507 &0.0504 &0.0527
  \\ \hline
\end{tabular}
\end{center}
\end{table}

\begin{table}[!htbp]
\scriptsize
\begin{center}
\caption{Type I error probabilities of the AGMCTs in the unknown variance case for designs that adapt the candidate dose-response models, as well as the corresponding tests in a non-adaptive design.}\label{sim: type I error, unknown variance}
\begin{tabular}{|l|l|l|l|}
\hline
\multicolumn{4}{|c|}{$N_1=N_2=60$}   \\ \hline
& T      & F      & N         \\ \hline
Adaptive& 0.0489& 0.0493 &0.0481   \\ \hline
Non-adaptive & 0.0523& 0.0533& 0.0532  \\ \hline

\multicolumn{4}{|c|}{$N_1=N_2=120$}   \\ \hline
& T      & F      & N          \\ \hline
Adaptive& 0.0470 &0.0466 &0.0459 \\ \hline
Non-adaptive & 0.0497& 0.0503 &0.0511 \\ \hline

\multicolumn{4}{|c|}{$N_1=N_2=180$}   \\ \hline
& T      & F      & N         \\ \hline
Adaptive& 0.0474 &0.0489 &0.0479  \\ \hline
Non-adaptive& 0.0458 &0.0478 &0.0473 \\ \hline

\multicolumn{4}{|c|}{$N_1=N_2=240$}   \\ \hline
& T      & F      & N        \\ \hline
Adaptive& 0.0481& 0.0490& 0.0482  \\ \hline
Non-adaptive&0.0507& 0.0476& 0.0479  \\ \hline
\end{tabular}
\end{center}
\end{table}
	

\begin{thebibliography}{9}
\bibitem{bib1} Bauer, P. and K\"{o}hne, K. (1994). Evaluation of experiments with adaptive interim analyses. \textit{Biometrics} \textbf{50}, 1029--1041.

\bibitem{bib2} Bauer, P. and R\"{o}hmel, J. (1995). An adaptive method for establishing a dose-response relationship. \textit{Statistics in Medicine} \textbf{14}, 1595--1607.

\bibitem{bib3} Bornkamp, B., Bretz, F., Dette, H., and Pinheiro, J. C. (2011). Response-adaptive dose-finding under model uncertainty. \textit{Annals of Applied Statistics} \textbf{5}, 1611--1631.

\bibitem{bib4} Brannath, W., Gutjahr, G., and Bauer, P. (2012). Probabilistic foundation of confirmatory adaptive designs. \textit{Journal of the American Statistical Association} \textbf{107}, 824--832.

\bibitem{bib5} Brannath, W., Koenig, F., and Bauer, P. (2007). Multiplicity and flexibility in clinical trials. \textit{Pharmaceutical Statistics} \textbf{6}, 205--216.

\bibitem{bib6} Brannath, W., Posch, M., and Bauer, P. (2002). Recursive combination tests. \textit{Journal of the American Statistical Association} \textbf{97}, 236--244.

\bibitem{bib7} Bretz, F., Pinheiro, J. C., and Branson, M. (2005). Combining multiple comparisons and modeling techniques in dose-response studies. \textit{Biometrics} \textbf{61}, 738--748.

\bibitem{bib8} Dragalin, V. (2006). Adaptive designs: Terminology and classification. \textit{Drug Information Journal} \textbf{40}, 425--435.

\bibitem{bib9} Dragalin, V., Bornkamp, B., Bretz, F., Miller, F., Padmanabhan, S. K., Patel, N., Perevozskaya, I., Pinheiro, J., and Smith, J. R. (2010). A simulation study to compare new adaptive dose-ranging designs. \textit{Statistics in Biopharmaceutical Research} \textbf{2}, 487--512.

\bibitem{bib10} Fisher, R. A. (1932). \textit{Statistical Methods for Research Workers}. Oliver and Boyd, London, UK.

\bibitem{bib11} Food and Drug Administration (2019). Adaptive Designs for Clinical Trials of Drugs and Biologics: Guidance for Industry. Food and Drug Administration, Washington DC, USA. https://www.fda.gov/media/78495/download

\bibitem{bib12} Franchetti, Y., Anderson, S. J., and Sampson, A. R. (2013). An adaptive two-stage dose-response design method for establishing proof of concept. \textit{Journal of Biopharmaceutical Statistics} \textbf{23}, 1124--1154.

\bibitem{bib13} Gaydos, B., Anderson, K. M., Berry, D., Burnham, N., Chuang-Stein, C., Dudinak, J., Fardipour, P., Gallo, P., Givens, S., Lewis, R., Maca, J., Pinheiro, J., Pritchett, Y., and Krams, M. (2009). Good practices for adaptive clinical trials in pharmaceutical product development. \textit{Drug Information Journal} \textbf{43}, 539--556.

\bibitem{bib14} Kost, J. T. and McDermott, M. P. (2002).  Combining dependent $P$-values. \textit{Statistics and Probability Letters} \textbf{60},  183--190.

\bibitem{bib15} Ma, S. and McDermott, M. P. (2020). Generalized multiple contrast tests in dose-response studies.  \textit{Statistics in Medicine} \textbf{39}, 757--772.

\bibitem{bib16} Mercier, F., Bornkamp, B., Ohlssen, D., and Wallstroem, E. (2015). Characterization of dose-response for count data using a generalized MCP-mod approach in an adaptive dose-ranging trial. \textit{Pharmaceutical Statistics} \textbf{14}, 359--367.

\bibitem{bib17} Miller, F. (2010). Adaptive dose-finding: Proof of concept with type I error control. \textit{Biometrical Journal} \textbf{52}, 577--589.

\bibitem{bib18} M\"{u}ller, H.-H. and Sch\"{a}fer, H. (2001). Adaptive group sequential designs for clinical trials: Combining the advantages of adaptive and of classical group sequential approaches. \textit{Biometrics} \textbf{57}, 886--891.

\bibitem{bib19} M\"{u}ller, H.-H. and Sch\"{a}fer, H. (2004). A general statistical principle for changing a design any time during the course of a trial. \textit{Statistics in Medicine} \textbf{23}, 2497--2508.

\bibitem{bib20} Pinheiro, J., Bornkamp, B., Glimm, E., and Bretz, F. (2014). Model-based dose finding under model uncertainty using general parametric models. \textit{Statistics in Medicine} \textbf{33}, 1646--1661.

\bibitem{bib21} Posch, M., Timmesfeld, N., K\"{o}nig, F., and M\"{u}ller, H. H. (2004). Conditional rejection probabilities of Student's t-test and design adaptations. \textit{Biometrical Journal} \textbf{46}, 389--403.

\bibitem{bib22} Robertson, T., Wright, F. T., and Dykstra, R. L. (1988). \textit{Order Restricted Statistical Inference}. Wiley, New York, NY.

\bibitem{bib23} Stouffer, S. A., Suchman, E. A., DeVinney, L. C., Star, S. A., and Williams, R. M. (1949). \textit{The American Soldier, Volume 1: Adjustment during Army Life}. Princeton University Press, Princeton, NJ.

\bibitem{bib24} Thomas, N. (2017). Understanding MCP-MOD dose finding as a method based on linear regression. \textit{Statistics in Medicine} \textbf{36}, 4401--4413.

\bibitem{bib25} Tippett, L. H. C. (1931).  \textit{The Method of Statistics}. Williams and Northgate, London, UK.

\bibitem{bib26} Wassmer, G. and Brannath, W. (2016). \textit{Group Sequential and Confirmatory Adaptive Designs in Clinical Trials}. Springer, Basel, Switzerland.

\end{thebibliography}
\end{document}